\documentclass[aps,twocolumn,floatfix,showpacs,amsmath,amsfonts]{revtex4}

\usepackage{graphicx}
\usepackage{bm}
\usepackage{units}
\usepackage{color}
\begin{document}

\title{Linewidths in excitonic absorption spectra of cuprous oxide}

\author{Frank Schweiner}
\author{J\"org Main}
%\email{main@itp1.uni-stuttgart.de}
\author{G\"unter Wunner}
%\email{wunner@itp1.uni-stuttgart.de}
\affiliation{Institut f\"ur Theoretische Physik 1, Universit\"at Stuttgart,
  70550 Stuttgart, Germany}
\date{\today}

\begin{abstract}
We present a theoretical calculation of the absorption spectrum of
cuprous oxide $\left(\mathrm{Cu_{2}O}\right)$ based on the general
theory developed by Y.~Toyozawa. An inclusion not only of acoustic
phonons but also of optical phonons and of specific properties of
the excitons in $\mathrm{Cu_{2}O}$ like the central-cell corrections
for the $1S$-exciton allows us to calculate the experimentally observed
line widths in experiments by T. Kazimierczuk \emph{et al}
[Nature \textbf{514}, 343, (2014)]
within the same order of magnitude, which demonstrates
a clear improvement in comparison to earlier work on this topic. We
also discuss a variety of further effects, which explain the still
observable discrepancy between theory and experiment but can
hardly be included in theoretical calculations.
\end{abstract}

\pacs{71.35.-y,~63.20.-e}

\maketitle

\section{Introduction}

Ever since the first formulation of their concept by Frenkel~\cite{SOK9_31F1,TOE_2,TOE_4},
Peierls~\cite{TOE_3}, and Wannier~\cite{TOE_5} in the 1930s, and
their experimental discovery in cuprous oxide $\left(\mathrm{Cu_{2}O}\right)$
by Gross and Karryjew in 1952~\cite{GRE_4}, excitons are of large
physical interest, because they are the quanta of the fundamental
optical excitations in both insulators and semiconductors in the visible
and ultraviolet spectrum of light. Excitons are so-called quasi-particles
being composed of an electron and a positivley charged hole. Wannier
excitons extend over many unit cells of the crystal and can be treated
within a very simple approach as an analogue of the hydrogen atom.
The corresponding Schr\"{o}dinger equation, which describes these excitons,
is the so-called Wannier equation~\cite{NM5_7,TOE,NM5_9}. 

Very recently, the hydrogen-like series could be observed
experimentally for the so-called yellow exciton in $\mathrm{Cu_{2}O}$ 
for the first time up to a large principal
quantum number of $n=25$~\cite{GRE}. 
This detection has brought new interest to the field of excitons~\cite{80,QC,28,FEM}.
However, the line widths detected in Ref.~\cite{GRE} 
differ from earlier theoretical calculations on this topic~\cite{2}, which leads 
us to a new investigation of the main parameters describing the shape of the
excitonic absorption lines.

The main features, which make $\mathrm{Cu_{2}O}$ one of the most
investigated semiconductors relating to excitons, are the large excitonic
binding energy of $R_{\mathrm{exc}}\approx86\,\mathrm{meV}$~\cite{80} and the
non-degeneracy of its uppermost valence band justifying the simple-band
model with a hydrogen-like exciton spectrum
\begin{equation}
E_{n\boldsymbol{K}}=E_{\mathrm{gap}}-\frac{R_{\mathrm{exc}}}{n^{2}}+
\frac{\hbar^{2}\boldsymbol{K}^{2}}{2M}.\label{eq:EnK}
\end{equation}
Besides the band gap energy $E_{\mathrm{gap}}$, we also include
the energy due to a finite momentum $\hbar\boldsymbol{K}$ of the
center of mass. By $M$ we denote the mass of the exciton in effective-mass
approximation. 
Beyond the simple-band model, one often has to account for a variety of further effects of the
solid. Possible corrections of this model include,
e.g., central-cell corrections~\cite{1}, a coupling of the
uppermost valence band to other valence bands~\cite{7,28}, and especially
the interaction with phonons, which are the quasi-particles of lattice
vibrations. This interaction is, besides the effect of impurities
in the crystal, the main cause for an asymmetric broadening and shifting
of the excitonic lines observed in absorption spectra~\cite{TOE}.
The general theory for the effect of phonons on excitonic spectra
has been developed by Toyozawa in the late 1950s and early 1960s~\cite{0,85,18,2}.

In the following we will apply the formulas of To\-yo\-zawa to the yellow
$nP$-excitons considering several corrections. This allows us to
calculate the observed line widths within the same order of 
magnitude when compared to the experiment~\cite{GRE}.
In Sec.~\ref{sec:23} we present the main results of Toyozawa's
theory.
\textcolor{black}{In contrast to earlier works on this topic, we perform calculations including 
all exciton states and no approximations as regards the phonon wave vector~\cite{2,73}.
In Sec.~\ref{sec:4} we calculate for the first time the effect of both acoustic phonons 
and optical phonons as well as the central-cell corrections of the $1S$-exciton 
state~\cite{1} on the line widths in the absorption spectrum.
Furthermore, we present a detailed list of a variety of effects explaining the 
remaining differences between theory and experiment in Sec.~\ref{sec:5}.}
Finally, we give a short summary and outlook in Sec.~\ref{sec:6}.

\section{Theory\label{sec:23}}

We will not present the complete theory of exciton-phonon coupling
here, but only present the main results of Toyozawa's theory. Readers
being interested in this topic are referred to Refs.~\cite{0,85,18,2,SST}.

In general, the exciton couples to two different kinds of phonons:
to longitudinal acoustic phonons (LA) via deformation potential coupling~\cite{M1_7}
and to longitudinal optical phonons (LO) via the Fr\"{o}hlich interaction~\cite{AP3}.
For both interactions the interaction Hamiltonian is of the same form
in second quantization
\begin{eqnarray}
H_{\mathrm{exc-ph}} & = & i\sum_{\boldsymbol{q}}\sum_{\nu\nu'\boldsymbol{K}}\lambda_{s}
\left(\boldsymbol{q},\,\nu\nu'\right)\nonumber \\
\nonumber \\
 &  & \times\left(a_{s}^{\phantom{\dagger}}\left(\boldsymbol{q}\right)-a_{s}^{\dagger}
 \left(-\boldsymbol{q}\right)\right)B_{\nu\boldsymbol{K}}^{\dagger}
 B_{\nu'\boldsymbol{K}-\boldsymbol{q}}^{\phantom{\dagger}}.
\end{eqnarray}
By $a_{s}^{\left(\dagger\right)}\left(\boldsymbol{q}\right)$ we denote
the operators annihilating (creating) a phonon in the mode $s\boldsymbol{q}$.
The operators $B_{\nu\boldsymbol{K}}^{\left(\dagger\right)}$ annihilate
(create) excitons with momentum $\hbar\boldsymbol{K}$ in the state
$\left|\nu\right\rangle =\left|nlm\right\rangle $. Since we make
use of the simple hydrogen-like model, we treat the quantum numbers
$n$, $l$ and $m$ as known from atomic physics as good quantum numbers;
although this is generally not the case due to the cubic symmetry
of the solid~\cite{28,80}. We will discuss this problem in Sec.~\ref{sec:5}.
The coupling matrix elements are given by
\begin{subequations}
\begin{equation}
\lambda_{\mathrm{LA}}\left(\boldsymbol{q},\,\nu\nu'\right)=f_{\mathrm{LA}}\left(q\right)
\left[D_{\mathrm{e}}q_{\mathrm{e}}\left(\boldsymbol{q},\,\nu\nu'\right)-
D_{\mathrm{h}}q_{\mathrm{h}}\left(\boldsymbol{q},\,\nu\nu'\right)\right]
\end{equation}
with
\begin{equation}
f_{\mathrm{LA}}\left(q\right)=\sqrt{\frac{\hbar}{2c_{\mathrm{LA}}\rho V}}q^{\frac{1}{2}}
\end{equation}
\end{subequations}
for LA phonons with the dispersion $\omega_{\mathrm{LA}}\left(\boldsymbol{q}\right)=c_{\mathrm{LA}}q$
including the velocity of sound $c_{\mathrm{LA}}$ and by
\begin{subequations}
\begin{equation}
\lambda_{\mathrm{LO}}\left(\boldsymbol{q},\,\nu\nu'\right)=f_{\mathrm{LO}}\left(q\right)
\left[q_{\mathrm{e}}\left(\boldsymbol{q},\,\nu\nu'\right)-q_{\mathrm{h}}\left(\boldsymbol{q},\,\nu\nu'\right)\right]
\end{equation}
with
\begin{equation}
f_{\mathrm{LO}}\left(q\right)=\sqrt{\frac{\hbar e^{2}\omega_{\mathrm{LO}}}{2V\varepsilon_{0}}
\left(\frac{1}{\varepsilon_{\mathrm{b}}}-\frac{1}{\varepsilon_{s}}\right)}\frac{1}{q}
\end{equation}
\end{subequations}
for LO phonons with the dispersion $\omega_{\mathrm{LO}}\left(\boldsymbol{q}\right)=\mathrm{const}$.
These matrix elements include the mass density $\rho$ and volume
$V$ of the solid, the deformation coupling potentials $D_{\mathrm{e}/\mathrm{h}}$
of conduction band and valence band, the dielectric constants above
$\left(\varepsilon_{\mathrm{b}}\right)$ and below $\left(\varepsilon_{\mathrm{s}}\right)$
the optical resonance as well as the effective charges as defined
by Toyozawa~\cite{0}:
\begin{subequations}
\begin{eqnarray}
q_{\mathrm{e}}\left(\boldsymbol{q},\,\nu\nu'\right) & = & \int\mathrm{d}\boldsymbol{r}\,
\psi_{\nu}^{*}\left(\boldsymbol{r}\right)\psi_{\nu'}^{\phantom{*}}\left(\boldsymbol{r}\right)
e^{i\frac{m_{\mathrm{h}}}{M}\boldsymbol{q}\boldsymbol{r}},\label{eq:qe}\\
\nonumber \\
q_{\mathrm{h}}\left(\boldsymbol{q},\,\nu\nu'\right) & = & \int\mathrm{d}\boldsymbol{r}\,
\psi_{\nu}^{*}\left(\boldsymbol{r}\right)\psi_{\nu'}^{\phantom{*}}\left(\boldsymbol{r}\right)
e^{-i\frac{m_{\mathrm{e}}}{M}\boldsymbol{q}\boldsymbol{r}}.\label{eq:qh}
\end{eqnarray}
\end{subequations}
By $m_{\mathrm{e/h}}$ we denote the effective masses of electron
and hole.

The interaction with phonons leads to peaks of asymmetric Loretzian
shape in the absorption spectrum. The absorption coefficient depending
on the frequency of light is given by~\cite{0,2}
\begin{widetext}
\begin{equation}
\alpha\left(\omega\right)=\sum_{\nu}\,\frac{\alpha_0}{\omega}\tilde{F}_{\nu}
\left(\omega\right)\frac{\hbar\tilde{\Gamma}_{\nu\boldsymbol{0}}\left(\omega\right)+
2\tilde{A}_{\nu}\left(\omega\right)\left[\hbar\omega-\tilde{E}_{\nu\boldsymbol{0}}
\left(\omega\right)\right]}{\left[\hbar\omega-\tilde{E}_{\nu\boldsymbol{0}}
\left(\omega\right)\right]^{2}+\left[\hbar\tilde{\Gamma}_{\nu\boldsymbol{0}}
\left(\omega\right)\right]^{2}}\label{eq:alpha}
\end{equation}
with the energy shift
\begin{equation}
\tilde{\Delta}_{\nu\boldsymbol{0}}\left(\omega\right)=\tilde{E}_{\nu\boldsymbol{0}}
\left(\omega\right)-E_{\nu\boldsymbol{0}}=\Delta_{\nu\nu\boldsymbol{0}}\left(\omega\right)
+\sum_{\nu'\neq\nu}\frac{\left|\Delta_{\nu\nu'\boldsymbol{0}}\left(\omega\right)\right|^{2}-
\left|\Gamma_{\nu\nu'\boldsymbol{0}}\left(\omega\right)\right|^{2}}
{E_{\nu\boldsymbol{0}}-E_{\nu'\boldsymbol{0}}},\label{eq:E}
\end{equation}
the line broadening
\begin{equation}
\tilde{\Gamma}_{\nu\boldsymbol{0}}\left(\omega\right)=\Gamma_{\nu\nu\boldsymbol{0}}
\left(\omega\right)+\sum_{\nu'\neq\nu}2\:\mathrm{Re}\left(\frac{\Delta_{\nu\nu'\boldsymbol{0}}
\left(\omega\right)\Gamma_{\nu'\nu\boldsymbol{0}}
\left(\omega\right)}{E_{\nu\boldsymbol{0}}-E_{\nu'\boldsymbol{0}}}\right),\label{eq:G}
\end{equation}
the scaling of the constant amplitude $\alpha_0$
\begin{eqnarray}
\tilde{F}_{\nu}\left(\omega\right) & = & \left|M_{\nu g}\right|^{2}+\sum_{\nu'\neq\nu}2\:\mathrm{Re}
\left(\frac{M_{\nu g}^{*}\Delta_{\nu\nu'\boldsymbol{0}}
\left(\omega\right)M_{\nu'g}^{\phantom{*}}}{E_{\nu\boldsymbol{0}}-E_{\nu'\boldsymbol{0}}}\right)\nonumber \\
\nonumber \\
 &  & +\sum_{\nu'\neq\nu}\sum_{\nu''\neq\nu}2\:\mathrm{Re}
 \left(\frac{M_{\nu g}^{*}\left(\Delta_{\nu\nu''\boldsymbol{0}}\left(\omega\right)
 \Delta_{\nu''\nu'\boldsymbol{0}}\left(\omega\right)-\Gamma_{\nu\nu''\boldsymbol{0}}
 \left(\omega\right)\Gamma_{\nu''\nu'\boldsymbol{0}}
 \left(\omega\right)\right)M_{\nu'g}}{\left(E_{\nu\boldsymbol{0}}-E_{\nu'\boldsymbol{0}}\right)
 \left(E_{\nu\boldsymbol{0}}-E_{\nu''\boldsymbol{0}}\right)}\right)\nonumber \\
\nonumber \\
 &  & +\sum_{\nu'\neq\nu}\sum_{\nu''\neq\nu}\mathrm{Re}\left(\frac{M_{\nu'g}^{*}
 \left(\Delta_{\nu'\nu\boldsymbol{0}}\left(\omega\right)\Delta_{\nu\nu''\boldsymbol{0}}
 \left(\omega\right)-\Gamma_{\nu'\nu\boldsymbol{0}}\left(\omega\right)
 \Gamma_{\nu\nu''\boldsymbol{0}}\left(\omega\right)\right)M_{\nu''g}}
 {\left(E_{\nu\boldsymbol{0}}-E_{\nu'\boldsymbol{0}}\right)
 \left(E_{\nu\boldsymbol{0}}-E_{\nu''\boldsymbol{0}}\right)}\right),\label{eq:F}
\end{eqnarray}
and the asymmetry $\tilde{A}_{\nu}\left(\omega\right)$, which can
be calculated from
\begin{eqnarray}
\tilde{A}_{\nu}\left(\omega\right)\tilde{F}_{\nu}\left(\omega\right) & = & 
\sum_{\nu'\neq\nu}\mathrm{Re}\left(\frac{M_{\nu g}^{*}\Gamma_{\nu\nu'\boldsymbol{0}}
\left(\omega\right)M_{\nu'g}^{\phantom{*}}}{E_{\nu\boldsymbol{0}}-E_{\nu'\boldsymbol{0}}}\right)\nonumber \\
\nonumber \\
 &  & +\sum_{\nu'\neq\nu}\sum_{\nu''\neq\nu}2\:\mathrm{Re}\left(\frac{M_{\nu g}^{*}
 \left(\Delta_{\nu\nu''\boldsymbol{0}}\left(\omega\right)\Gamma_{\nu''\nu'\boldsymbol{0}}
 \left(\omega\right)+\Gamma_{\nu\nu''\boldsymbol{0}}\left(\omega\right)\Delta_{\nu''\nu'\boldsymbol{0}}
 \left(\omega\right)\right)M_{\nu'g}}{\left(E_{\nu\boldsymbol{0}}-E_{\nu'\boldsymbol{0}}\right)
 \left(E_{\nu\boldsymbol{0}}-E_{\nu''\boldsymbol{0}}\right)}\right)\nonumber \\
\nonumber \\
 &  & +\sum_{\nu'\neq\nu}\sum_{\nu''\neq\nu}\mathrm{Re}\left(\frac{M_{\nu'g}^{*}
 \left(\Delta_{\nu'\nu\boldsymbol{0}}\left(\omega\right)\Gamma_{\nu\nu''\boldsymbol{0}}
 \left(\omega\right)-\Gamma_{\nu'\nu\boldsymbol{0}}\left(\omega\right)
 \Delta_{\nu\nu''\boldsymbol{0}}\left(\omega\right)\right)M_{\nu''g}}{\left(E_{\nu\boldsymbol{0}}
 -E_{\nu'\boldsymbol{0}}\right)\left(E_{\nu\boldsymbol{0}}-E_{\nu''\boldsymbol{0}}\right)}\right).\label{eq:A}
\end{eqnarray}

The quantity $M_{\nu g}$ denotes the transition matrix element between
the ground state $\left|0\right\rangle $ of the solid and the exciton
state $\left|\nu\right\rangle $ with $\boldsymbol{K}=0$ due to the
electron-photon interaction. In cuprous oxide the transition is parity-forbidden,
which results in~\cite{TOE}
\begin{equation}
M_{\nu g}=c\:\frac{n^{2}-1}{n^{5}}\delta_{l,1}\delta_{m,0}.\label{eq:Mnug}
\end{equation}
Since in both equations~(\ref{eq:F}) and~(\ref{eq:A}) $M_{\nu g}$
appears quadratically, the asymmetry $\tilde{A}_{\nu}\left(\omega\right)$
will be independent of the proportionality constant $c$. The main
difficulty in the implementation of the formulas given above is the
calculation of the quantities~\cite{0,2}

\begin{eqnarray}
\Gamma_{\nu_{2}\nu_{1}\boldsymbol{0}}\left(\omega\right) & = & \sum_{s\boldsymbol{q}}
\sum_{\nu_{3}}\frac{\pi}{\hbar}\lambda_{s}^{*}\left(\boldsymbol{q},\,\nu_{3}\nu_{2}\right)
\lambda_{s}^{\phantom{*}}\left(\boldsymbol{q},\,\nu_{3}\nu_{1}\right)\nonumber \\
\nonumber \\
 &  & \times\left[\left(n_{s}\left(\boldsymbol{q},\, T\right)+1\right)\delta
 \left(E_{\nu_{3}\boldsymbol{q}}+\hbar\omega_{s}\left(\boldsymbol{q}\right)-\hbar\omega\right)
 +n_{s}\left(\boldsymbol{q},\, T\right)\delta\left(E_{\nu_{3}\boldsymbol{q}}-
 \hbar\omega_{s}\left(\boldsymbol{q}\right)-\hbar\omega\right)\right],\label{eq:gamma}
\end{eqnarray}
and
\begin{eqnarray}
\Delta_{\nu_{2}\nu_{1}\boldsymbol{0}}\left(\omega\right) & = & \sum_{s\boldsymbol{q}}
\sum_{\nu_{3}}\lambda_{s}^{*}\left(\boldsymbol{q},\,\nu_{3}\nu_{2}\right)\lambda_{s}^{\phantom{*}}
\left(\boldsymbol{q},\,\nu_{3}\nu_{1}\right)\nonumber \\
\nonumber \\
 &  & \times\left[\left(n_{s}\left(\boldsymbol{q},\, T\right)+1\right)\mathcal{P}
 \left(\frac{1}{\hbar\omega-E_{\nu_{3}\boldsymbol{q}}-\hbar\omega_{s}
 \left(\boldsymbol{q}\right)}\right)+n_{s}\left(\boldsymbol{q},\, T\right)
 \mathcal{P}\left(\frac{1}{\hbar\omega-E_{\nu_{3}\boldsymbol{q}}+\hbar\omega_{s}
 \left(\boldsymbol{q}\right)}\right)\right].\label{eq:delta}
\end{eqnarray}
The symbol $\mathcal{P}$ denotes the principal value. We can write
\begin{eqnarray}
\mathcal{P}\left(\frac{1}{x}\right) =\mathcal{P}\int\mathrm{d}E\,\frac{1}{E}
\delta\left(E-x\right)\lim_{\epsilon\rightarrow0^{+}}\left(\int_{-\infty}^{-\epsilon}
\mathrm{d}E\,\frac{1}{E}\delta\left(E-x\right)+\int_{\epsilon}^{\infty}\mathrm{d}E\,
\frac{1}{E}\delta\left(E-x\right)\right).\label{eq:princ}
\end{eqnarray}
\end{widetext}
The average thermal occupation of phononic states at a temperature
$T$ is given by~\cite{SST}
\begin{equation}
n_{s}\left(\boldsymbol{q},\, T\right)=\frac{1}{e^{\hbar\omega_{s}
\left(\boldsymbol{q}\right)/k_{\mathrm{B}}T}-1}.\label{eq:ns}
\end{equation}

The evaluation of $\Gamma_{\nu_{2}\nu_{1}\boldsymbol{0}}\left(\omega\right)$
and $\Delta_{\nu_{2}\nu_{1}\boldsymbol{0}}\left(\omega\right)$ as well as their application to
$\mathrm{Cu_{2}O}$ are presented in the Appendix.

\section{Results and Discussion\label{sec:45}}

\subsection{Contributions to the line widths\label{sec:4}}

In the following we will discuss the different contributions to the
line widths $\tilde{\Gamma}_{\nu\boldsymbol{0}}\left(\omega\right)$ in Eq.~(\ref{eq:G})
for $\mathrm{Cu_{2}O}$ at the very low temperature of $T=1.2\,\mathrm{K}$~\cite{GRE}.
The relevant material parameters are listed in Table \ref{tab:Material-parameters-of}.
Although the unit cell of $\mathrm{Cu_{2}O}$ comprises $6$ atoms,
which amounts in $15$ optical phonon modes, there are only two LO-phonon
modes with $\Gamma_{4}^{-}$-symmetry contributing to the Fr\"{o}hlich
interaction~\cite{1}.

\begin{table}
\begin{centering}

\caption{Material parameters of $\mathrm{Cu_{2}O}$ used in our calculations.
$m_{0}$ denotes the free electron mass. All values are taken from
Ref.~\cite{SOK1_82L1} unless otherwise stated. \label{tab:Material-parameters-of}}

\begin{tabular}{lll}
\hline 
Parameter & $\phantom{\int_{\int}^{\int}}$ & Value\tabularnewline
\hline 
Lattice constant & $\phantom{\int_{\int}^{\int}}$ & $a=4.27\times10^{-10}\,\mathrm{m}$\tabularnewline
Mass density & $\phantom{\int_{\int}^{\int}}$ & $\rho=6.09\,\mathrm{\frac{g}{cm^{3}}}$\tabularnewline
Band gap energy & $\phantom{\int_{\int}^{\int}}$ & $E_{\mathrm{g}}=2.17\,\mathrm{eV}$\tabularnewline
Effective masses~\cite{JPC76} & $\phantom{\int_{\int}^{\int}}$ & $m_{\mathrm{e}}=0.99m_{0}$\tabularnewline
 & $\phantom{\int_{\int}^{\int}}$ & $m_{\mathrm{h}}=0.58m_{0}$\tabularnewline
Dielectric constants & $\phantom{\int_{\int}^{\int}}$ & $\varepsilon_{\mathrm{s1}}=7.5$,~$\varepsilon_{\mathrm{b1}}=7.11$\tabularnewline
 & $\phantom{\int_{\int}^{\int}}$ & $\varepsilon_{\mathrm{s2}}=7.11$,~$\varepsilon_{\mathrm{b2}}=6.46$\tabularnewline
Sound wave velocity & $\phantom{\int_{\int}^{\int}}$ & $c_{\mathrm{LA}}=4.5405\times10^{3}\,\mathrm{\frac{m}{s}}$\tabularnewline
Energy of $\Gamma_{4}^{-}$-LO phonons~\cite{1} & $\phantom{\int_{\int}^{\int}}$ & $\hbar\omega_{\mathrm{LO,\,1}}=18.7\,\mathrm{meV}$\tabularnewline
 & $\phantom{\int_{\int}^{\int}}$ & $\hbar\omega_{\mathrm{LO,\,2}}=87\,\mathrm{meV}$\tabularnewline
Deformation potentials~\cite{M1} & $\phantom{\int_{\int}^{\int}}$ & $D_{\mathrm{e}}=2.4\,\mathrm{eV}$\tabularnewline
 & $\phantom{\int_{\int}^{\int}}$ & $D_{\mathrm{h}}=2.2\,\mathrm{eV}$\tabularnewline
Rydberg energy~\cite{80} & $\phantom{\int_{\int}^{\int}}$ & $R_{\mathrm{exc}}=86\,\mathrm{meV}$\tabularnewline
\hline 
\end{tabular}
\par\end{centering}

\end{table}

For our discussion we will especially consider the line parameters
of the $2P$-exciton since it has always been wondered which effects
lead to the large broadening of this line~\cite{0,15,GRE_11_5,GRE_11_6}. 
We discuss the contributions to these parameters in several steps.

\emph{Step 1}: We start with the most simple case, in which we neglect the optical
phonons, set the frequency $\omega$ to $E_{\nu\boldsymbol{0}}/\hbar$
and neglect the so-called intraband-contributions~\cite{0}, i.e.,
we only include those parts of Eqs.~(\ref{eq:E})-(\ref{eq:A}),
which do not contain sums over $\nu'$.
The approximation of setting $\omega\approx E_{\nu\boldsymbol{0}}/\hbar$
is justified since $\tilde{\Gamma}_{\nu\boldsymbol{0}}\left(\omega\right)$
is a slowly varying function with $\omega$ \cite{2, 15}. The formula
(\ref{eq:gamma}) includes a sum over all excitonic states. In order
to calculate the quantity $\Gamma_{\nu_{2}\nu_{1}\boldsymbol{0}}\left(\omega\right)$
within reasonable time, we have to restrict the infinite sum to a
finite one via
\begin{equation}
\sum_{\nu_{3}}\rightarrow\sum_{n_{3}=1}^{n_{\mathrm{max}}}\;\sum_{l_{3}=0}^{n_{3}-1}\;
\sum_{m_{3}=-l_{3}}^{l_{3}}\label{eq:sum}
\end{equation}
with $n_{\mathrm{max}}\leq7$. As it has also been done by Toyozawa~\cite{0},
one may at first include only states having the same principal quantum
number as the one considered. This means for the $2P$-exciton that
the sum reads
\begin{equation}
\sum_{\nu_{3}}\rightarrow\sum_{n_{3}=2}^{2}\;\sum_{l_{3}=0}^{1}\;\sum_{m_{3}=-l_{3}}^{l_{3}}.
\end{equation}
This yields very small values for the line width and the energy shift
\begin{subequations}
\begin{eqnarray}
\tilde{\Gamma}_{210\,\boldsymbol{0}}\left(E_{210\,\boldsymbol{0}}/\hbar\right) & \approx & 1.70\times10^{-9}\,\mathrm{eV},\\
\nonumber \\
\tilde{\Delta}_{210\,\boldsymbol{0}}\left(E_{210\,\boldsymbol{0}}/\hbar\right) & \approx & -9.72\times10^{-6}\,\mathrm{eV}.
\end{eqnarray}
\end{subequations}

\begin{figure}

\includegraphics[width=1.0\columnwidth]{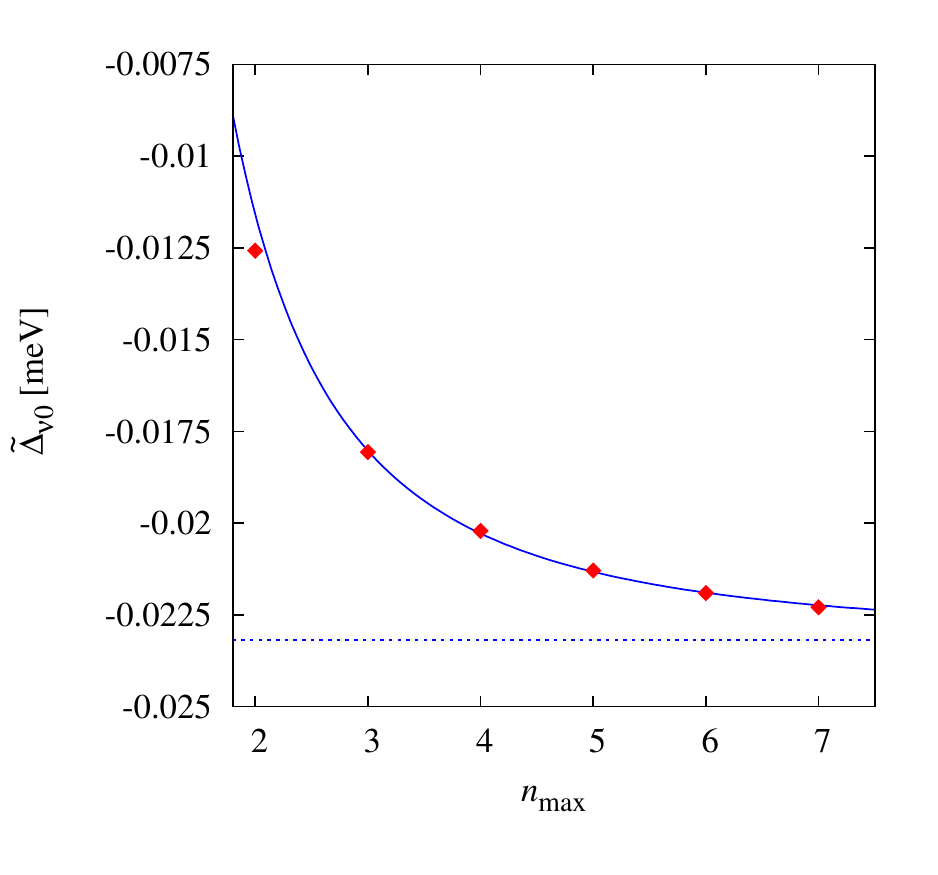}

\caption{(Color online) In order to evaluate the quantities 
$\Gamma_{\nu_{2}\nu_{1}\boldsymbol{0}}\left(\omega\right)$
and $\Delta_{\nu_{2}\nu_{1}\boldsymbol{0}}\left(\omega\right)$, one
has to cut the infinite sums over $\nu$ in the formulas at a finite
value $n_{\mathrm{max}}$ of the principal quantum number $n$ (cf.
Eq.~(\ref{eq:sum})). Here we show the values obtained for $\tilde{\Delta}_{210\,\boldsymbol{0}}$
in dependence on $n_{\mathrm{max}}$ for \emph{Step 2}. The final value 
$\tilde{\Delta}_{210\,\boldsymbol{0}}=-2.32\times10^{-5}\,\mathrm{eV}$ (dashed line)
is then calculated from an extrapolation. We used $f\left(n_{\mathrm{max}}\right)=a/n_{\mathrm{max}}^{2}+b$
as a fitting function for $n_{\mathrm{max}}\geq3$ (solid line). \label{fig:In-order-to}}
\end{figure}

\emph{Step 2}: An obviously better approach is to evaluate the complete sum (\ref{eq:sum})
with different $n_{\mathrm{max}}$ and ex\-tra\-po\-late the values obtained
for $\tilde{\Gamma}_{210\,\boldsymbol{0}}$ and $\tilde{\Delta}_{210\,\boldsymbol{0}}$
to the final value for $n_{\mathrm{max}}\rightarrow\infty$. To this
aim we fit a function of the form $f\left(n_{\mathrm{max}}\right)=a/n_{\mathrm{max}}^{2}+b$
to our values. We depict this procedure in Fig.~\ref{fig:In-order-to}.
This approach yields
\begin{subequations}
\begin{eqnarray}
\tilde{\Gamma}_{210\,\boldsymbol{0}}\left(E_{210\,\boldsymbol{0}}/\hbar\right) & \approx & 9.87\times10^{-7}\,\mathrm{eV},\\
\nonumber \\
\tilde{\Delta}_{210\,\boldsymbol{0}}\left(E_{210\,\boldsymbol{0}}/\hbar\right) & \approx & -2.32\times10^{-5}\,\mathrm{eV}.
\end{eqnarray}
\end{subequations}
This already shows that the $1S$-exciton state has a large influence
on the line width of the $2P$-state.

\emph{Step 3}: At very low temperatures only few LA phonons are thermally excited.
We therefore expect the optical phonons to increase the line width
considerably; especially since the energy of one of these phonons
$\left(\hbar\omega_{\mathrm{LO,\,1}}=18.7\,\mathrm{meV}\right)$ is
of the same magnitude as the energetic difference between two exciton
states $\left(E_{210\,\boldsymbol{0}}-E_{410\,\boldsymbol{0}}\approx17.25\,\mathrm{meV}\right)$.
Including optical phonons, we obtain
\begin{subequations}
\begin{eqnarray}
\tilde{\Gamma}_{210\,\boldsymbol{0}}\left(E_{210\,\boldsymbol{0}}/\hbar\right) & \approx & 3.45\times10^{-5}\,\mathrm{eV},\\
\nonumber \\
\tilde{\Delta}_{210\,\boldsymbol{0}}\left(E_{210\,\boldsymbol{0}}/\hbar\right) & \approx & -8.39\times10^{-3}\,\mathrm{eV}.
\end{eqnarray}
\end{subequations}

\emph{Step 4}: Up to now we have assumed that the line width 
$\tilde{\Gamma}_{210\,\boldsymbol{0}}\left(\omega\right)$
is a slowly varying function of the frequency of light.
For this reason we have set $\omega\approx E_{\nu\boldsymbol{0}}/\hbar$.
In the literature it has been discussed that it is necessary to account
for the frequency dependence in order to describe the asymmetry of
the lines correctly~\cite{15}. On the other hand, Toyozawa already
stated in Ref.~\cite{2} that the line shape would not be of asymmetric
Lorentzian shape if $\tilde{\Gamma}_{210\,\boldsymbol{0}}\left(\omega\right)$
varied strongly with $\omega$. We see that the energy shift $\tilde{\Delta}_{210\,\boldsymbol{0}}$
is several $\mathrm{meV}$ large. Since the absorption peak is centered
around $\tilde{E}_{\nu\boldsymbol{0}}\left(\omega\right)$, we evaluate
the line parameters within the range $\omega_{\mathrm{min}}\leq\omega\leq\omega_{\mathrm{max}}$
with
\begin{equation}
\hbar\omega_{\mathrm{min}}=E_{210\,\boldsymbol{0}}
-2\left|\tilde{\Delta}_{210\,\boldsymbol{0}}\left(E_{210\,\boldsymbol{0}}/\hbar\right)\right|
\end{equation}
and
\begin{equation}
\hbar\omega_{\mathrm{max}}=E_{210\,\boldsymbol{0}}
\end{equation}
to determine their frequency dependence. It is found that 
$\tilde{\Gamma}_{210\,\boldsymbol{0}}\left(\omega\right)$
increases slowly with $\omega$ while $\tilde{\Delta}_{210\,\boldsymbol{0}}\left(\omega\right)$
decreases strongly :
\begin{subequations}
\begin{eqnarray}
\tilde{\Gamma}_{210\,\boldsymbol{0}}\left(\omega_{\mathrm{min}}\right) & \approx & 3.30\times10^{-5}\,\mathrm{eV},\\
\tilde{\Gamma}_{210\,\boldsymbol{0}}\left(\omega_{\mathrm{max}}\right) & \approx & 3.45\times10^{-5}\,\mathrm{eV},\\
\nonumber \\
\tilde{\Delta}_{210\,\boldsymbol{0}}\left(\omega_{\mathrm{min}}\right) & \approx & -6.97\times10^{-3}\,\mathrm{eV},\\
\tilde{\Delta}_{210\,\boldsymbol{0}}\left(\omega_{\mathrm{max}}\right) & \approx & -8.39\times10^{-3}\,\mathrm{eV}.
\end{eqnarray}
\end{subequations}
The effect on the line width may be more important in external fields,
which would mix different excitonic states~\cite{0,23}.

%\begin{widetext}
\begin{figure*}

\includegraphics[width=1.9\columnwidth]{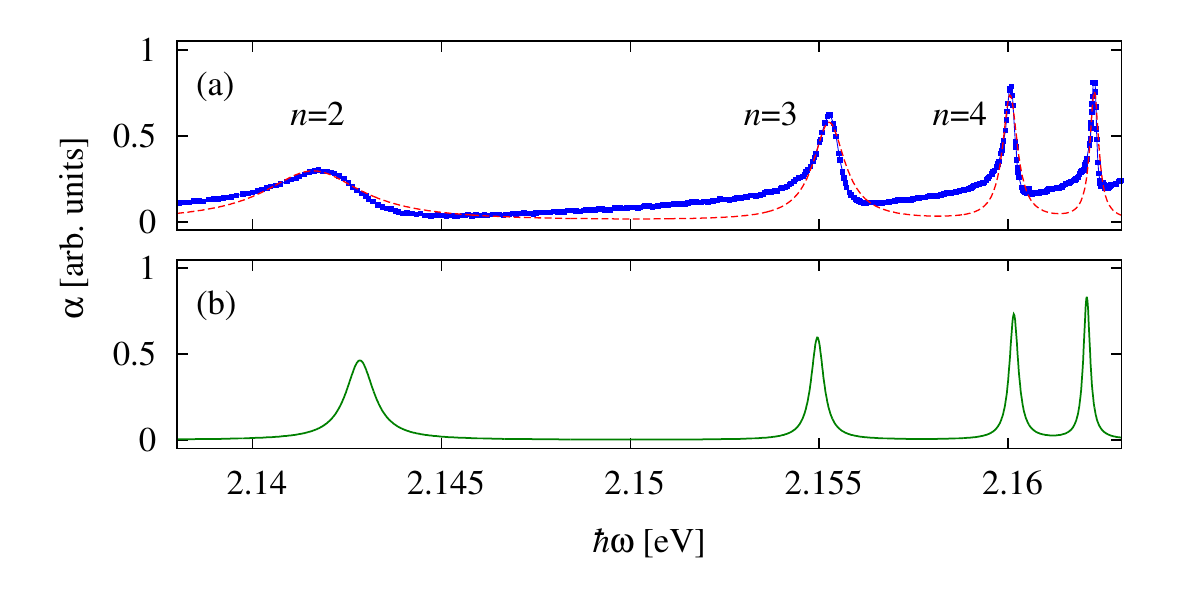}

\caption{(Color online) Comparison of (a) the experimental spectrum~\cite{GRE} 
with (b) the calculated line shapes 
using Eq.~(\ref{eq:alpha}) and the values listed in Table~\ref{tab:The-final-values}.
Since we do not know the proportionality constant $c$ in Eq.~(\ref{eq:Mnug}),
we chose arbitrary values for the amplitudes $\tilde{F}_{\nu}\left(\omega\right)$.
We shifted the experimental values by an amount of $-6~\mathrm{meV}$ for a better comparison.
\textcolor{black}{The experimental values have been fitted by Lorentzians to obtain 
the experimental line widths (red dashed line).}
\label{fig:Comparison-of-the}}

\end{figure*}
%\end{widetext}

\emph{Step 5}: An important effect concerns the $1S$-exciton of the yellow series
of $\mathrm{Cu_{2}O}$. The mean distance between electron and hole
is so small that this exciton can hardly be treated as a Wannier exciton.
The corrections that have to be made due to this small distance are
known as the central-cell corrections. They lead to a higher mass
of the $1S$-exciton of $\tilde{M}\approx3m_{0}$ and to a smaller
excitonic Bohr radius of $\tilde{a}_{\mathrm{exc}}\approx0.53\,\mathrm{nm}$~\cite{1}.
These corrections are now included in the excitonic wave function
$\psi_{100}$ and in the excitonic energies 
\begin{equation}
E_{100\,\boldsymbol{K}}=E_{\mathrm{gap}}-
\frac{\tilde{R}_{\mathrm{exc}}}{n^{2}}+\frac{\hbar^{2}\boldsymbol{K}^{2}}{2\tilde{M}}.
\end{equation}
The binding energy $\tilde{R}_{\mathrm{exc}}\approx153\,\mathrm{meV}$
of the $1S$-exciton differs much from the excitonic Rydberg constant of the rest of
the yellow exciton series. The central-cell corrections have a significant
influence on the line width and increase it by a factor of about~$17$
to
\begin{subequations}
\begin{eqnarray}
\tilde{\Gamma}_{210\,\boldsymbol{0}}\left(\omega_{\mathrm{min}}\right) & \approx & 6.12\times10^{-4}\,\mathrm{eV},\\
\tilde{\Gamma}_{210\,\boldsymbol{0}}\left(\omega_{\mathrm{max}}\right) & \approx & 5.53\times10^{-4}\,\mathrm{eV},\\
\nonumber \\
\tilde{\Delta}_{210\,\boldsymbol{0}}\left(\omega_{\mathrm{min}}\right) & \approx & -6.98\times10^{-3}\,\mathrm{eV},\\
\tilde{\Delta}_{210\,\boldsymbol{0}}\left(\omega_{\mathrm{max}}\right) & \approx & -8.18\times10^{-3}\,\mathrm{eV}.
\end{eqnarray}
\end{subequations}

\emph{Step 6}: We now investigate the influence of intraband scattering. Therefore,
we also consider the sums of the form $\sum_{\nu'\neq\nu}$ in Eqs.~(\ref{eq:E})-(\ref{eq:A}), where we also
cut these sums at the same value of $n_{\mathrm{max}}$. In contrast
to the expectation of Toyozawa~\cite{2}, the effect of this type of
scattering on the line width is quite small. We obtain
\begin{subequations}
\begin{eqnarray}
\tilde{\Gamma}_{210\,\boldsymbol{0}}\left(\omega_{\mathrm{min}}\right) & \approx & 4.04\times10^{-4}\,\mathrm{eV},\\
\tilde{\Gamma}_{210\,\boldsymbol{0}}\left(\omega_{\mathrm{max}}\right) & \approx & 4.94\times10^{-4}\,\mathrm{eV},\\
\nonumber \\
\tilde{\Delta}_{210\,\boldsymbol{0}}\left(\omega_{\mathrm{min}}\right) & \approx & -7.14\times10^{-3}\,\mathrm{eV},\\
\tilde{\Delta}_{210\,\boldsymbol{0}}\left(\omega_{\mathrm{max}}\right) & \approx & -8.57\times10^{-3}\,\mathrm{eV}.
\end{eqnarray}
Nevertheless, the asymmetry of the lines can be explained only by
intraband scattering. The value of 
\begin{eqnarray}
\tilde{A}_{210}\left(\omega_{\mathrm{min}}\right) & \approx & -3.67\times10^{-2},\\
\tilde{A}_{210}\left(\omega_{\mathrm{max}}\right) & \approx & -3.69\times10^{-2},
\end{eqnarray}
\end{subequations}
is, however, very small in comparison with the large asymmetry of the
lines observed in~\cite{GRE}. \textcolor{black}{We will discuss this discrepancy in Sec.~\ref{sec:5}.}

\emph{Step 7}: In the literature a large asymmetry has also been assigned to a coupling
of the bound exciton states to the continuum states~\cite{15,15_9a,15_9b,26},
whose energies are given by
\begin{equation}
E_{\boldsymbol{k}\,\boldsymbol{K}}=E_{\mathrm{gap}}+\frac{\hbar^{2}\boldsymbol{k}^{2}}{2\mu}+
\frac{\hbar^{2}\boldsymbol{K}^{2}}{2M}\label{eq:EkK}
\end{equation}
in analogy to the hydrogen atom. However, an effect of the continuum
states can be excluded via a simple calculation: For the average occupation
of the phonon modes one obtains $n_{\mathrm{LO,\,1}}\left(\boldsymbol{q},\, T\right)=0$
for $T\lesssim25\,\mathrm{K}$ and $n_{\mathrm{LO,\,2}}\left(\boldsymbol{q},\, T\right)=0$
for $T\lesssim100\,\mathrm{K}$, i.e., only scattering processes with
the emission of phonons can take place at $T=1.2\,\mathrm{K}$. Furthermore,
the emission process can only take place if the arguments of the
$\delta$-functions in Eqs.~(\ref{eq:gamma}) and~(\ref{eq:delta})
are positive. This means for acoustic phonons
\begin{subequations}
\begin{equation}
\frac{\hbar^{2}\boldsymbol{k}^{2}}{2\mu}<\frac{1}{2}Mc_{\mathrm{LA}}^{2}-\frac{R_{\mathrm{exc}}}{n^{2}}
\end{equation}
and for optical phonons
\begin{equation}
\frac{\hbar^{2}\boldsymbol{k}^{2}}{2\mu}<-\hbar\omega_{\mathrm{LO}}-\frac{R_{\mathrm{exc}}}{n^{2}}.
\end{equation}
\end{subequations}
Therefore, only LA phonons play a role and only for the line shapes
of excitons with $n>32$. A contribution of the continuum states is
therefore impossible.

The final results including all of the corrections discussed above
are listed in Table \ref{tab:The-final-values}. \textcolor{black}{We also list the experimental line widths, 
which have been obtained by fitting Lorentzians 
or Elliotts formula to the experimental absorption spectrum 
(cf.~Fig.~\ref{fig:Comparison-of-the}).}
It can be seen
that we obtain the correct behavior of the line parameters with increasing
principal quantum number: The line widths decrease with increasing quantum number.

In Fig.~\ref{fig:Comparison-of-the} we compare the predicted line shapes
with the measured ones. It is obvious that our calculation cannot 
reproduce quantitatively the large asymmetry. 
However, the line widths differ only by a factor of~$\sim3.5$ or even~$\sim1.3$, 
which means that they are of the same order of magnitude. 
The observable difference in the position
of the lines can be explained on the one hand by small inaccuracies 
of the material constants used, 
on the other hand in terms of the complex valence band
structure of $\mathrm{Cu_{2}O}$. These facts
and further possible reasons for deviations from the experimental
spectrum will be discussed in Sec.~\ref{sec:5}. 

\textcolor{black}{A quantitative comparison of the calculated line widths with the results of previous 
works is not possible. In Ref.~\cite{0} it is reported that the calculated line 
width of the $2P$-exciton is several times smaller than the experimentally observed 
one but no value is given. In Ref.~\cite{73} the calculated line widths are several 
times larger than the experimental ones indicating the inappropriateness of the many 
approximations in that publication.}

\begin{table}
\begin{centering}

\caption{The final values for the line widths $\tilde{\Gamma}_{\nu\boldsymbol{0}}$,
the energy shifts $\tilde{\Delta}_{\nu\boldsymbol{0}}$
including all of the corrections discussed in Sec.~\ref{sec:4}.
The values are given at $\hbar\omega=E_{n10\,\boldsymbol{0}}-
\tilde{\Delta}_{n10\,\boldsymbol{0}}\left(E_{n10\,\boldsymbol{0}}/\hbar\right)$ 
(cf.~\emph{Step 4} of Sec.~\ref{sec:4}).
\textcolor{black}{In the last column the experimental line widths are listed~\cite{GRE}.}
\label{tab:The-final-values}}

\begin{tabular}{cccc}
\hline 
$\nu$ & $\tilde{\Gamma}_{\nu\boldsymbol{0}}\,\left[\mathrm{meV}\right]$ & 
$\tilde{\Delta}_{\nu\boldsymbol{0}}\,\left[\mathrm{meV}\right]$ & $\tilde{\Gamma}_{\nu\boldsymbol{0}}$(exp)$\,\left[\mathrm{meV}\right]$\tabularnewline
\hline 
$210$ & $0.453$ & $-7.737$ & $1.581$\tabularnewline
$310$ & $0.201$ & $-7.574$ & $0.511$\tabularnewline
$410$ & $0.144$ & $-6.551$ & $0.237$\tabularnewline
$510$ & $0.108$ & $-6.560$ & $0.142$\tabularnewline
\hline 
\end{tabular}
\par\end{centering}

\end{table}

\subsection{Further discussion\label{sec:5}}

In the above calculation we made some assumptions, which are discussed
in the following. We also discuss possible causes for a further
broadening of the lines, which may be difficult to be considered
in theory.

We have assumed that the dispersion of LA phonons is linear according to
$\omega_{\mathrm{LA}}=c_{\mathrm{LA}}q$ and that the dispersion of
LO phonons is constant. If we perform the $q$-integration according
to Eq.~(\ref{eq:intq}) only up to a value of $q_{\mathrm{max}}<q_{\mathrm{D}}$,
our results do not change for $\frac{1}{2}q_{\mathrm{D}}<q_{\mathrm{max}}<q_{\mathrm{D}}$,
i.e., we can always set the upper boundary of the integral to $\frac{1}{2}q_{\mathrm{D}}$.
Since the assumption of the LA dispersion relation to be linear in
$q$ holds for $q<\frac{1}{2}q_{\mathrm{D}}$~\cite{10a, 10a_12a,10a_12b,10a_12c},
its usage is retroactively justified. Furthermore, the change of the
energy of the LO phonons within this limit is very small~\cite{13}.

We have treated $l$ and $m$ as good quantum numbers in the above
calculations. This is in general not the case due to the cubic symmetry
of the crystal. Nevertheless, since $\mathrm{O_{h}}$ is the point
group with the highest symmetry, it may be justified to treat $l$
approximatively as a good quantum number~\cite{28,80}. However, one
would still have to calculate the correct linear combinations of states
with different $m$ quantum number, which then transform according
to the irreducible representations of the cubic group $\mathrm{O_{h}}$~\cite{28}.
This has not been done since we expect no effect from this rearrangement
of states.

\textcolor{black}{The asymmetry of the lines calculated in~\emph{Step 6} are considerably
smaller than the experimental values.
The large asymmetries can be explained in terms of 
Fano resonances and phonon replicas.
Phonon replicas describe, in particular for luminescence, 
the scattering of a polariton from the exciton-like branch 
of its dispersion relation to the photon-like branch with 
the simultaneous emission of LO phonons, or more simply the 
decay of an exciton with the emission of one photon and LO phonons.
In luminescence spectra the line shape then shows a square-root-like 
energy dependence due to the exciton density of states.
While LO-phonon replicas appear on the low-energy side in luminescence spectra, 
they can also appear on the high-energy side in absorption spectra~\cite{SO}.
In the case of $\mathrm{Cu_2 O}$ the $\Gamma_{3}^{-}$ LO phonon assists the $1S$ 
exciton formation and causes the square-root-like frequency dependence of the 
absorption coefficient, on which then absorption of the other exciton resonances 
is superimposed (see, e.g., Refs.~\cite{14,72,73}).
Since the transition amplitudes interfere destructively or constructively on the lower 
or higher energy side 
of the resonance with the continuum of the $\Gamma_{3}^{-}$ LO phonon, one obtains 
asymmetric line shapes of the exciton resonances in accordance with the theory of Fano resonances~\cite{71}.
Since the formulas of Sec.~\ref{sec:23} do not account for the phononic background, 
we could not determine correct asymmetries of the lines.
Note that the phononic background has been subtracted from the results of Ref.~\cite{GRE};
an absorption spectrum including it can be found, e.g., in Ref.~\cite{GRE_11}.
For further information on this topic, see also Refs.~\cite{UDPS,SO} and further references therein.}

The Rydberg energies $R_{\mathrm{exc}}$ of excitonic spectra are generally 
obtained from fits to experimental results.
Therefore, the value of $R_{\mathrm{exc}}$ for the yellow series of $\mathrm{Cu_{2}O}$ varies between
$86\,\mathrm{meV}$~\cite{80} and $97.2\,\mathrm{meV}$~\cite{12_23} in the literature. 
The same argument holds for the band gap energy $E_{\mathrm{gap}}$.
One reason for the deviations in the line positions in Fig.~\ref{fig:Comparison-of-the} 
is thus the uncertainity in these constants.
 
We have also assumed that the simple band model holds. Indeed, the results
in~\cite{GRE} show that this approximation is reasonable; but one
could also include the complete valence band structure in the theory~\cite{7,28}.
This makes an investigation of line widths almost impossible since
the energies $E_{\nu\boldsymbol{K}}$ have to be determined first
of all, and a separation of relative motion and the motion of the center
of mass is not possible~\cite{17,24}. The calculations
in Ref.~\cite{17} on the line widths of the $1S$-exciton states of
different semiconductors already show the main problems if one would
have to extend the theory to principal quantum numbers of $n\geq2$.
On the other hand, an inclusion of the complete valence band structure
results in a coupling of the yellow and green exciton series, especially
to the green $1S$-exciton state~\cite{7}. Since we found out that
the yellow $1S$-exciton state has a significant influence on the
line width of the $2P$-exciton state, we expect that the coupling
to the green $1S$-exciton state will lead to a further broadening
of this line. The coupling to the (energetically higher located) green
series may also be a reason for the large degree of asymmetry of the
lines.

The complex valence band structure is sometimes treated in a simple
approach in terms of quantum defects~\cite{80,GRE}. However, the results
of Ref.~\cite{80} show that this approach works well only at high
quantum numbers $\left(n\geq7\right)$. Therefore, we did not consider
quantum defects in our calculations.

The complex valence band structure also facilitates a coupling of
excitons to TA phonons~\cite{17,17_13,SST}. However, the effect of
TA phonons is reported to be half as large as the effect of LA phonons~\cite{10a,10a_28},
which is already very small in our case. The coupling to TA phonons
may be more important if external strains are applied to the crystal~\cite{PRB73}.
 
Impurities, especially point defects, in the crystal can lead to
a broadening of exciton line widths~\cite{12}. The effect of an increase
in the defect concentration has, according to Toyozawa~\cite{0}, the
same effect as a raising of the temperature. However, it has been
discussed in the literature that a large concentration of impurities will
lead to a more Gaussian or Voigt line shape~\cite{16,16_7,17}. This
cannot be seen in the line spectrum measured in Ref.~\cite{GRE},
for which reason we have to assume that the concentration of defects
is low. Certainly, one could also estimate the concentration of defects
experimentally by an extrapolation of the line width to $T\rightarrow0\,\mathrm{K}$~\cite{PRB58}. 
Furthermore, the effect of a movement of defects
being caused by phonons is said to be negligible~\cite{17}.

The Fr\"{o}hlich coupling constant is defined as~\cite{SST}
\begin{equation}
\phantom{space}\alpha^{\mathrm{F}}=\frac{e^{2}}{8\pi\varepsilon_{0}\hbar
\omega_{\mathrm{LO}}}\left(\frac{2M\omega_{\mathrm{LO}}}{\hbar}\right)^{\frac{1}{2}}
\left(\frac{1}{\varepsilon_{\mathrm{b}}}-\frac{1}{\varepsilon_{s}}\right)
\end{equation}
For $\mathrm{Cu_{2}O}$ we obtain $\alpha_{1}^{\mathrm{F}}\approx0.24$
and $\alpha_{2}^{\mathrm{F}}\approx0.20$. Since these values are
clearly smaller than $1$, we can neglect polaron corrections to the
energy and the mass of the excitons~\cite{SST,23}. 

In the unit cell of $\mathrm{Cu_{2}O}$ there are always four copper
atoms arranged in tetragonal symmetry~\cite{JPCS27}, but only in every
second tetragon an oxygen atom is located at its center. Since the
oxygen atoms are very small, there is a chance that there are sometimes
more than two oxygen atom in one unit cell. The excess atoms will
then occupy the free positions in the lattice and act as acceptors.
This results in small charges and in small internal electric fields,
which will influence the exciton and lead to a line broadening. However,
it is hard to account for these fields in theory.

The coupling between excitons and phonons is linear, i.e., there is
always only one phonon being involved in a scattering process. In the
literature, multi-phonon processes are said to be important in connection
with piezoelectric coupling~\cite{15}. Sometimes, they are even said
to be negligible~\cite{16}. \textcolor{black}{Since piezoelectric coupling 
is symmetry-forbidden in $\mathrm{Cu_{2}O}$, we do not consider multi-phonon processes.}

In general, there are no excitons in crystals but polaritons due to
the coupling to light~\cite{TOE}. In other materials than $\mathrm{Cu_{2}O}$
the excitonic $1S$-ground state is often dipole allowed. The resulting
large polariton coupling mainly changes the contribution of LA phonons
to the line widths but the contribution of the LO phonons only weakly~(see
Ref.~\cite{23} and further references therein). Since the LA-phononic
contribution is small for $\mathrm{Cu_{2}O}$, we expect
that the polariton effect will not change our results significantly, 
so that it can even be neglected~\cite{10a}.

We have shown that the central-cell corrections have a major
influence on the line width of the $2P$-exciton state. Besides the
central-cell corrections, which lead to an increase in the mass of
the $1S$-exciton, there exists also a $\boldsymbol{K}$-dependent
exchange interaction, which results in a $\boldsymbol{K}$-dependent
effective mass $\tilde{M}\left(\boldsymbol{K}\right)$ of this exciton~\cite{8}.
We expect the influence of the $\boldsymbol{K}$-dependency of the
mass $\tilde{M}$ to be small for the following reason: We have proven
that the effect of interband coupling on the line width is unimportant.
For this reason the main contribution to the line widths comes from
the region with $\boldsymbol{K}\approx\boldsymbol{0}$ and it is sufficient
to take the value $\tilde{M}\left(\boldsymbol{0}\right)$~(cf. the
illustrations of intraband and interband scattering in Ref.~\cite{TOE}).

\section{Summary and outlook\label{sec:6}}

We have calculated the main parameters describing the shape of the
excitonic absorption lines for the yellow exciton series of $\mathrm{Cu_{2}O}$
and compared our results to the experimentally observed lines of Ref.~\cite{GRE}.
Especially the calculated line width for yellow $2P$-exciton lies
within the same order of magnitude as the experimental one and differs
only by a factor of~$\sim3.5$, which is a significant improvement on
the result of Ref.~\cite{2}. Furthermore, we have discussed possible
reasons for the large broadening and the large asymmetry of the lines.
Of course, some of these special properties of $\mathrm{Cu_{2}O}$
could eventually be included in theory, but only with huge effort.

Recently, it has been shown that the yellow excitonic
line spectrum of $\mathrm{Cu_{2}O}$ in an external magnetic field
shows GUE statistics~\cite{QC}. This line statistics has been explained in terms
of the exciton-phonon coupling in the crystal. Therefore, it will be worthwhile
to extend our calculations by including a magnetic field in order to prove the 
GUE statistics theoretically.

\acknowledgments
We thank Ch.~Uihlein and D.~Fr\"{o}hlich for helpful discussions.

\appendix*
\section{Evaluation of \textmd{$\Gamma_{\nu_{2}\nu_{1}\boldsymbol{0}}\left(\omega\right)$}
and \textmd{$\Delta_{\nu_{2}\nu_{1}\boldsymbol{0}}\left(\omega\right)$}}

\textcolor{black}{We now present the evaluation of Eqs.~(\ref{eq:gamma}) and~(\ref{eq:delta}) as well as their application to $\mathrm{Cu_{2}O}$.}

Due to periodic boundary conditions, the values of the phononic wave
vector $\boldsymbol{q}$ are generally discrete~\cite{SST}. If we
apply the continuum approximation, in which the number of atoms $N$
of the solid goes to infinity and the lattice constant $a_{\mathrm{lat}}$
between the atoms goes to zero while the ratio $Na_{\mathrm{lat}}^{3}=V$
is kept constant, we can treat $\boldsymbol{q}$ as a continuous quantity
and replace the corresponding sums by integrals:
\begin{subequations}
\begin{equation}
\sum_{\boldsymbol{q}} \rightarrow \frac{V}{\left(2\pi\right)^{3}}\int\mathrm{d} \boldsymbol{q}
\end{equation}
with
\begin{equation}
\int\mathrm{d} \boldsymbol{q} = \int_{0}^{q_{\mathrm{D}}}\mathrm{d}q\, q^{2}
 \int_{0}^{\pi}\mathrm{d}q_{\vartheta}\,\sin q_{\vartheta}\int_{0}^{2\pi}
 \mathrm{d}q_{\varphi}.\label{eq:intq}
\end{equation}
\end{subequations}
The upper boundary $q_{\mathrm{D}}$ of the $q$-integral is given
by the boundary of the first Brillouin zone (BZ) and can be calculated
from the Debye model~\cite{SST}.
\textcolor{black}{In order to evaluate the integral over $\boldsymbol{q}$, the dependence
of the effective charges on the angles $q_{\vartheta}$ and $q_{\varphi}$
has to be determined.} To this end we substitute the variable $\boldsymbol{r}$
in the integrals of Eqs.~(\ref{eq:qe}) and~(\ref{eq:qh}) by 
$\boldsymbol{u}=\boldsymbol{A}^{\mathrm{T}}\boldsymbol{r}$
with a rotation matrix $\boldsymbol{A}$, for which 
$\boldsymbol{A}^{\mathrm{T}}\boldsymbol{q}=q\hat{\boldsymbol{e}}_{z}$
holds. By $\hat{\boldsymbol{e}}_{z}$ we denote the unit vector in
$z$-direction. If we denote by $\boldsymbol{R}_{\hat{\boldsymbol{n}}\varphi}$
the rotation matrix describing the rotation about an axis $\hat{\boldsymbol{n}}$
by an angle $\varphi$, we can express $\boldsymbol{A}$ as
\begin{equation}
\boldsymbol{A}=\boldsymbol{R}_{\hat{\boldsymbol{e}}_{z}\left(-q_{\alpha}\right)}
\boldsymbol{R}_{\hat{\boldsymbol{e}}_{y}\left(-q_{\vartheta}\right)}
\boldsymbol{R}_{\hat{\boldsymbol{e}}_{z}\left(-q_{\varphi}\right)}
\end{equation}
with an arbitrary angle $q_{\alpha}$. The hydrogen-like wave functions
$\psi_{\nu}$ of the exciton read 
\begin{equation}
\psi_{\nu}\left(\boldsymbol{r}\right)=R_{nl}(r)Y_{lm}(\vartheta,\,\varphi)
\end{equation}
with the spherical harmonics $Y_{lm}(\vartheta,\,\varphi)$.
For the radial part $R_{nl}(r)$ we take the well-known
functions of the hydrogen atom~\cite{Messiah1}, but replace the Bohr
radius $a_{0}$ by the excitonic Bohr radius $a_{\mathrm{exc}}$,
which is given by~\cite{SST}
\begin{equation}
a_{\mathrm{exc}}=a_{0}\frac{\mathrm{Ry}}{\varepsilon_{\mathrm{s1}}
R_{\mathrm{exc}}}\approx1.116\,\mathrm{nm},
\end{equation}
with the Rydberg energy $\mathrm{Ry}$ and the dielectric constant
$\varepsilon_{\mathrm{s1}}$, which is given together with all other material 
parameters of $\mathrm{Cu_2 O}$ in Table~\ref{tab:Material-parameters-of}. 

After the substitution, we make use of the special properties of the
spherical harmonics under rotations~\cite{ED}:
\begin{eqnarray}
\psi_{\nu}\left(\boldsymbol{A}\boldsymbol{u}\right) & = & e^{-\frac{i}{\hbar}
q_{\alpha}\hat{\boldsymbol{e}}_{z}\boldsymbol{L}}e^{-\frac{i}{\hbar}q_{\vartheta}
\hat{\boldsymbol{e}}_{y}\boldsymbol{L}}e^{-\frac{i}{\hbar}q_{\varphi}\hat{\boldsymbol{e}}_{z}
\boldsymbol{L}}\psi_{\nu}\left(\boldsymbol{u}\right)\nonumber \\
\nonumber \\
 & = & \mathcal{D}\left(q_{\alpha},\, q_{\vartheta},\, q_{\varphi}\right)
 \psi_{\nu}\left(\boldsymbol{u}\right)\nonumber \\
\nonumber \\
 & = & R_{nl}\left(u\right)\sum_{m'=-l}^{l}Y_{lm'}\left(u_{\vartheta},\, u_{\varphi}\right)\nonumber \\
 & & \qquad\qquad\qquad\times D_{m'm}^{l}\left(q_{\alpha},\, q_{\vartheta},\, q_{\varphi}\right).
\end{eqnarray}
The complex factors $D_{m'm}^{l}\left(q_{\alpha},\, q_{\vartheta},\, q_{\varphi}\right)$
are the matrix elements of the operator 
$\mathcal{D}\left(q_{\alpha},\, q_{\vartheta},\, q_{\varphi}\right)$
corresponding to the spherical harmonics, i.e., 
\begin{equation}
D_{m'm}^{l}\left(q_{\alpha},\, q_{\vartheta},\, q_{\varphi}\right)=\left\langle lm'
\left|\mathcal{D}\left(q_{\alpha},\, q_{\vartheta},\, q_{\varphi}\right)\right|lm\right\rangle .
\end{equation}
Since the final expressions do not depend
on $q_{\alpha}$, it is possible to include an additional integral
$\frac{1}{2\pi}\int\mathrm{d}q_{\alpha}$. \textcolor{black}{Making
use of the properties of the matrices $D_{m'm}^{l}$~\cite{TOE}, we can easily 
evaluate the integrals over $q_{\vartheta}$ and $q_{\varphi}$.}
%\begin{widetext}
%\begin{eqnarray}
%\Lambda_{s}\left(q,\nu_{3}\nu_{2}\nu_{1}\right) & = & \sum_{f=\max
%\left(\left|l_{3}-l_{2}\right|,\left|l_{3}-l_{1}\right|\right)}^{\min\left(l_{3}+l_{2},l_{3}+l_{1}\right)}\;
%\sum_{g=-\min\left(l_{3},l_{2}\right)}^{\max\left(l_{3},l_{2}\right)}\;
%\sum_{h=-\min\left(l_{3},l_{1}\right)}^{\max\left(l_{3},l_{1}\right)}4\pi
%\left(2f+1\right)\delta_{m_{1}m_{2}}\nonumber \\
%\nonumber \\
% &  & \times\left(\begin{array}{ccc}
%l_{3} & l_{1} & f\\
%g & h & -g-h
%\end{array}\right)\left(\begin{array}{ccc}
%l_{3} & l_{2} & f\\
%h & g & -h-g
%\end{array}\right)\left(\begin{array}{ccc}
%l_{3} & l_{1} & f\\
%m_{3} & m_{1} & -m_{3}-m_{1}
%\end{array}\right)\left(\begin{array}{ccc}
%l_{3} & l_{2} & f\\
%m_{3} & m_{1} & -m_{3}-m_{1}
%\end{array}\right)\nonumber \\
%\nonumber \\
% &  & \times\left(D_{\mathrm{e}}^{s}\left\langle n_{2}l_{2}g\left|
%e^{-i\frac{m_{\mathrm{h}}}{M}qz}\right|n_{3}l_{3}g\right\rangle -
%D_{\mathrm{h}}^{s}\left\langle n_{2}l_{2}g\left|e^{i\frac{m_{\mathrm{e}}}{M}qz}
%\right|n_{3}l_{3}g\right\rangle \right)\nonumber \\
%\nonumber \\
% &  & \times\left(D_{\mathrm{e}}^{s}\left\langle n_{1}l_{1}h\left|
%e^{-i\frac{m_{\mathrm{h}}}{M}qz}\right|n_{3}l_{3}h\right\rangle -
%D_{\mathrm{h}}^{s}\left\langle n_{1}l_{1}h\left|e^{i\frac{m_{\mathrm{e}}}{M}qz}
%\right|n_{3}l_{3}h\right\rangle \right)^{*}.
%\end{eqnarray}
%\end{widetext}
\textcolor{black}{The arising matrix elements of the form
\begin{eqnarray}
\left\langle nlm\left|e^{iaz}\right|n'l'm\right\rangle  & = & 
\int\mathrm{d}\boldsymbol{r}\, R_{nl}(r)R_{n'l'}(r)e^{iar\cos\vartheta}\nonumber \\
\nonumber \\
 &  & \times Y_{lm}^{*}(\vartheta,\,\varphi)Y_{l'm}^{\phantom{*}}(\vartheta,\,\varphi)\label{eq:matrixel}
\end{eqnarray}
are calculated using Mathematica. }

The evaluation of the integral over $q$ is straightforward.
At first, we interchange the integral over $q$ with the integral
belonging to the principal value in Eq.~(\ref{eq:delta}). Then we
treat the arguments of the delta functions in Eqs.~(\ref{eq:gamma})
and~(\ref{eq:delta}) as functions of $q$ and use the relation
\begin{equation}
\delta\left(f\left(q\right)\right)=\sum_{i}\left|\left.\frac{\partial f}
{\partial q}\right|_{q=q_{i}}\right|^{-1}\delta\left(q-q_{i}\right),\label{eq:delta-1}
\end{equation}
where the sum is over all roots $q_{i}$ of $f\left(q\right)$. 

The final task is the evaluation of the integral with the pricipal
value in $\Delta_{\nu\nu'\boldsymbol{0}}\left(\omega\right)$. This
will be done numerically using Hartree units. One can read from the
delta functions obtained by using Eq.~(\ref{eq:delta-1}) for which
energies $E$ there will be a contribution to the integral. According
to the values of the material parameters of $\mathrm{Cu_{2}O}$ the
maximum and minimum value of $E$ are given by
\begin{subequations} 
\begin{eqnarray}
E_{\mathrm{max}} & = & R_{\mathrm{exc}}+\hbar\omega_{\mathrm{LO,\, max}}>0,\\
\nonumber \\
E_{\mathrm{min}} & = & -E_{\mathrm{max}}-\frac{\hbar^{2}q_{\mathrm{max}}^{2}}{2M}<0,
\end{eqnarray}
\end{subequations}
where $\hbar\omega_{\mathrm{LO,\, max}}$ denotes the energy of
the LO-phonon mode with highest energy. Since $\left|E_{\mathrm{min}}\right|>E_{\mathrm{max}}$
holds, we can replace the upper value of the integral by $-E_{\mathrm{min}}$
and rewrite the principal value integral as an improper integral
\begin{equation}
\mathcal{P}\int_{E_{\mathrm{min}}}^{-E_{\mathrm{min}}}\mathrm{d}E\, f\left(E\right)=
\lim_{\epsilon\rightarrow0}\int_{\epsilon}^{-E_{\mathrm{min}}}\mathrm{d}E\,
\left(f\left(E\right)+f\left(-E\right)\right),
\end{equation}
which is then evaluated using Gaussian quadrature and a standard algorithm
for improper integrals.

%\rule[0.5ex]{0.9\columnwidth}{1pt}

\end{document}